\documentclass{parcfd}

\usepackage{nopageno}
\usepackage{graphicx}
\usepackage{amsmath}
\usepackage{amsfonts}
\usepackage{amssymb}
\usepackage{hyperref}
\usepackage{layout}
\usepackage{floatrow}
\newfloatcommand{capbtabbox}{table}[][] 
\usepackage{blindtext}

\title{WaterLily.jl: A differentiable fluid simulator in Julia with fast heterogeneous execution}

\author{Gabriel D. Weymouth$^{*}$, Bernat Font$^{\dag}$}

\heading{Gabriel D. Weymouth, Bernat Font}

\address{$^{*}$ Technical University Delft \\
Mechanical, Maritime \& Materials Engineering (3mE)\\
Delft, the Netherlands\\
G.D.Weymouth@tudelf.nl, https://weymouth.github.io
\and
$^{\dag}$Barcelona Supercomputing Center\\
Computer Applications in Science and Engineering (CASE)\\
Barcelona, Spain\\
bernat.font@bsc.es, https://b-fg.github.io}


\abstract{Integrating computational fluid dynamics (CFD) software into optimization and machine-learning frameworks is hampered by the rigidity of classic computational languages and the slow performance of more flexible high-level languages. WaterLily.jl is an open-source incompressible viscous flow solver written in the Julia language. The small code base is multi-dimensional, multi-platform and backend-agnostic (serial CPU, multi-threaded, \& GPU execution). The simulator is differentiable and uses automatic-differentiation internally to immerse solid geometries and optimize the pressure solver. The computational time per time step scales linearly with the number of degrees of freedom on CPUs, and we see up to a 182x speed-up using CUDA kernels. This leads to comparable performance with Fortran solvers on many research-scale problems opening up exciting possible future applications on the cutting edge of machine-learning research.
}

\begin{document}

\section{Introduction}

During the last decade, the computational fluid dynamics (CFD) community has embraced the surge of machine learning (ML) and the new developments in hardware architecture, such as general-purpose GPUs. Hence, classic CFD solvers based on low-level programming languages (C, Fortran) and CPU memory-distributed computations are now adapted to accommodate these new tools.

On one hand, the integration of high-level ML libraries and low-level CFD solvers is not straight-forward, aka. the two-language problem. When deploying a ML model online with the CFD solver, data exchange is often performed at disk level, significantly slowing down the overall runtime because of disk i/o operations. An improved way to exchange data is performed through memory, either using Unix sockets \cite{Rabault2019, Font2021} or an in-memory distributed database \cite{Kurz2022}, which increases the software complexity. On the other hand, porting classic CFD solvers to GPU is also a non-trivial task which often requires the input and expertise of GPU vendors \cite{Romero2022}.

Julia \cite{Bezanson2017} is an open-source, compiled, dynamic, and composable programming language specifically designed for scientific computing which can help tackle such software challenges. High-level libraries and low level code can co-exist without compromising computing performance. Moreover, its excellent meta-programming capabilities, dynamic types, and multiple-dispatch strategy maximizes code re-usability. A great example of this is the KernelAbstractions.jl library \cite{KernelAbstractions.jl}, which enables writing heterogeneous kernels for different backends (multithreaded CPU, NVIDIA, AMD, and others) in a single framework. Julia has been also tested in many HPC systems, and the reader is referred to \cite{JuliaHPC} for a comprehensive review.

\section{Computational approach}

WaterLily solves the incompressible Navier-Stokes equations on a uniform Cartesian background grid using a third-order finite volume scheme with explicit adaptive time stepping and implicit Large Eddy Simulation turbulence modelling \cite{Margolin2006}. The Boundary Data Immersion Method \cite{Weymouth2011JCP} which accurately enforces general boundary conditions through modification of the pressure Poisson matrix \cite{Lauber2022} is used. The pressure system is solved using Geometric-MultiGrid (GMG), with a solution time that scales linearly with the number of cells \cite{WEYMOUTH2022105620}.

Julia's flexible and fast programming capabilities enabled the implementation of WaterLily to have many special features. For example, Automatic Differentiation (AD) to define all of the properties of the immersed geometry from a user-defined signed-distance function and coordinate mapping function. Indeed, the entire solver is differentiable, which has been used to develop accelerated data-driven GMG methods \cite{WEYMOUTH2022105620}. However, the most important Julia features for implementing the solver to run on heterogeneous back-ends are (i) the typing system, (ii) the meta-programming capabilities, and (iii) the rich open-source packages. Multiple-dispatch enables simple functions (such as array-scaling or reduction) to be written at high-level by the user and the compiler will specialize the code for efficient execution on the CPU or GPU. For more specialized tasks, meta-programming is used to generate efficient code based on a general kernel. As an example, the gradient of the \texttt{n}-dimensional pressure field \texttt{p} is applied to the velocity field \texttt{u} using
\begin{verbatim}
for i in 1:n  # apply pressure gradient
    @loop u[I,i] -= coeff[I,i]*(p[I]-p[I-del(i)]) over I in inside(p)
end
del(i) = ntuple(j -> j==i ? 1 : 0,n)
\end{verbatim}
where \texttt{del(i)} defines a step in direction \texttt{i}. \texttt{@loop} is a macro which evaluates this kernel at the points \texttt{I} inside the field using the KernelAbstractactions.jl package \cite{KernelAbstractions.jl} to generate optimized code for each back-end. This macro is used for nearly every loop in the code-base, enabling the efficient heterogeneous flow solver to be written in only around 800 lines of code!

Note that there are drawbacks to this simple kernel-driven approach. For one - each kernel tends to be fairly short. While this makes the code easy to read, it is not great for loading up the GPU or CPU threads. In the example above, the n-loops \textit{can} be combined into a single loop, improving the loading. However, this isn't possible for many loops in the code, and would require significant refactoring for many others, meaning only large array operations will see a speed up. The second drawback is that kernels cannot be used for loops that must iterate through points in order, such as in many relaxation methods. Because of this, we now use Conjugate-Gradient smoothing within the GMG levels.

\section{Applications and results}

A performance comparison between baseline serial execution, multi-threaded CPU execution, and GPU execution is presented next. The baseline execution does not make use of the KernelAbstractions.jl library, whereas the parallel executions do. Two different 3D cases are considered: the Taylor--Green vortex (TGV) at $Re=10^5$ and flow past a donut at $Re=10^3$ (see Figure 2). A total of 0.1 convective time units is simulated for each case. The main difference between the cases is the presence of a solid boundary, which makes the pressure solver dominate the execution time of the simulation. Otherwise, the convection-diffusion routine becomes the most expensive. Different grid sizes are considered for each case. For the TGV, $N=(2^p, 2^p, 2^p)$ where $p=\{5, 6, 7, 8\}$ is considered. For the donut case, $N=(2^{p+1}, 2^p, 2^p)$ where $p=\{4, 5, 6, 7\}$ is considered. The grids are selected so that the tests fit in a NVIDIA GeForce GTX 1650 Ti GPU card. The CPU execution is performed on an Intel Core i7-10750H x6 processor, with thread-affinity set to each physical core (following, multi-threaded CPU execution is denoted as ``CPU" and serial CPU execution as ``serial").

The TGV results are presented first. As observed in Figure 1, the runtime of the time-stepping routine \texttt{simstep!} increases linearly with the grid size in the serial execution. The CPU multi-threaded execution and the GPU execution significantly speed up this runtime. Specifically, it can be observed in Table 1 that the CPU and GPU executions can outperform the serial execution by a factor of 9 and 70 respectively. Table 2 breaks down the main routines in \texttt{simstep!}, showing prominent speed-ups for the convection-diffusion routine \texttt{convdiff!} and pressure solver \texttt{project!}, which are the most expensive kernels in the incompressible flow solver. The \texttt{convdiff!} routine dominates the TGV test case, and so the speed-ups of around 10x on CPU and 70x on GPU are reflected in the overall simulation speed up. The donut test is dominated by \texttt{project!} and sees up to 23x speed-up on CPU and 182x on GPUs. Note that the CPU execution stagnates at $log_2(N)=21$ whereas the GPU execution still improves up to the finest grid.

\newpage

\begin{figure}[h!]
\begin{floatrow}
\ffigbox{%
  \includegraphics[width=0.9\linewidth]{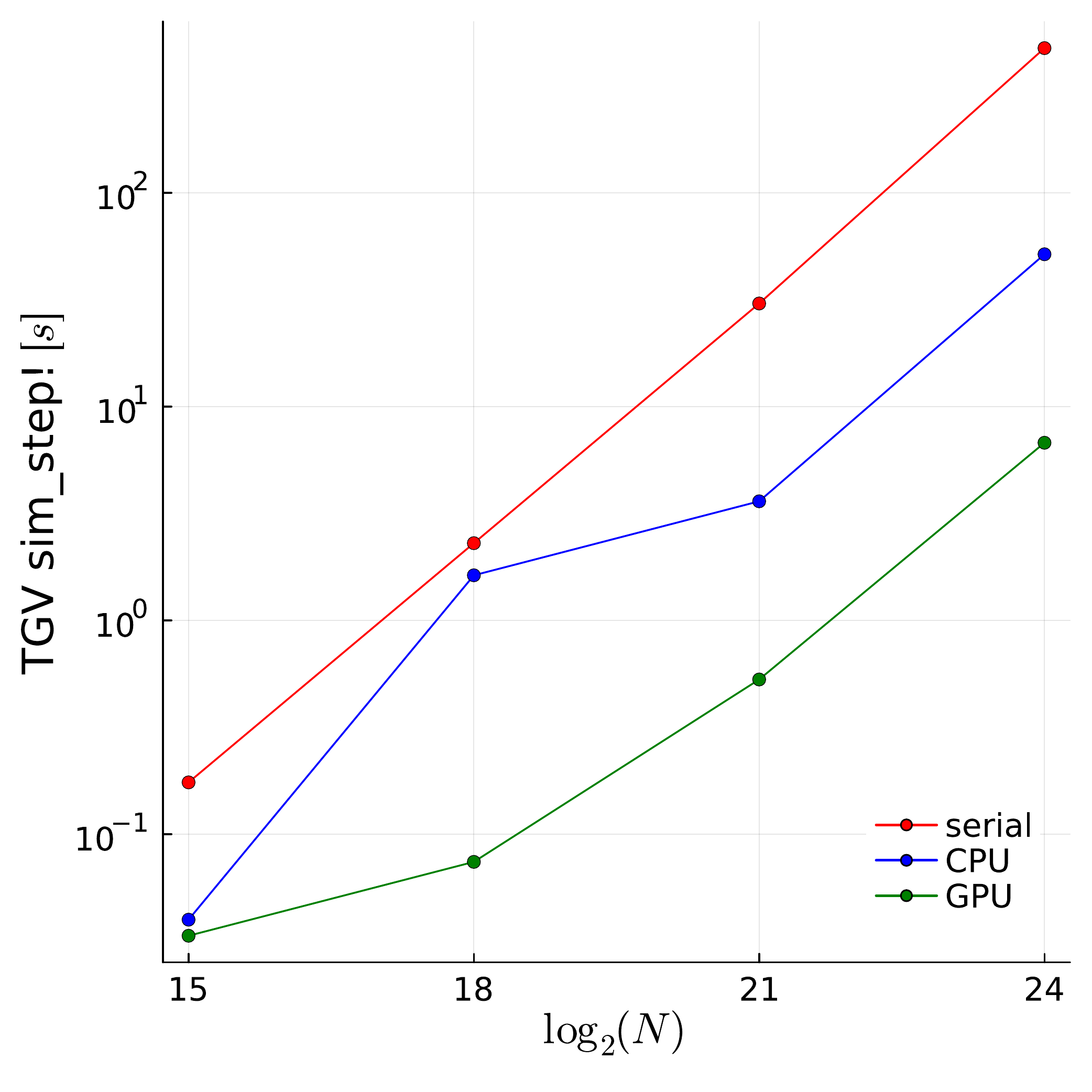}
  \label{fig:TGV_benchmark}
}{%
  \caption{TGV execution time for 0.1 convective time units on the different backends and grids.}%
}
\ffigbox{%
  \includegraphics[width=1\linewidth]{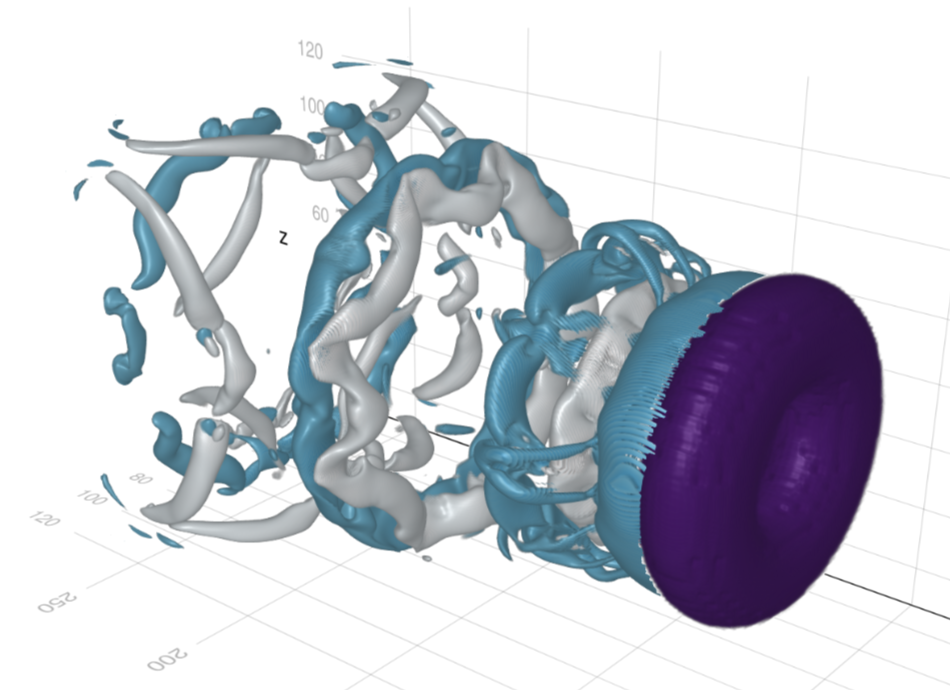}
  \label{fig:donut}
}{%
  \caption{Vorticity azimuth component isosurface of the donut test case.}%
}
\end{floatrow}
\end{figure}

\begin{figure}[h!]
\begin{floatrow}
\capbtabbox{%
  \begin{tabular}{crr} \hline
  $log_2(N)$ & CPU & GPU \\ \hline
  & TGV & \\ \hline
  15 & 4.39 & 5.22 \\
  18 & 1.41 & 30.95\\
  21 & 8.42 & 57.35\\
  24 & 9.21 & 70.10\\ \hline
  & Donut & \\ \hline
  12 &  3.51 & 2.53\\
  15 & 20.95 & 29.93 \\
  18 & 25.32 & 100.15\\
  21 & 23.57 & 181.97\\ \hline
  \end{tabular}\label{tab:simstep_benchmark}
}{%
  \caption{Speed-ups obtained with respect to the baseline serial execution for the different test cases and grids.}%
}
\capbtabbox{%
  \begin{tabular}{ccrr} \hline
  $log_2(N)$ & kernel & CPU & GPU \\ \hline
  21 & \texttt{convdiff!} & 8.76 & 65.10 \\
  21 & \texttt{BDIM!} & 18.39 & 97.17 \\
  21 & \texttt{BC!} & 1.95 & 7.47 \\
  21 & \texttt{project!} & 16.44 & 105.93 \\
  21 & \texttt{CFL!} & 5.52 & 23.67 \\ \hline
  24 & \texttt{convdiff!} & 9.38 & 73.61 \\
  24 & \texttt{BDIM!} & 17.54 & 94.44 \\
  24 & \texttt{BC!} & 2.38 & 10.09 \\
  24 & \texttt{project!} & 16.05 & 111.48 \\ 
  24 & \texttt{CFL} & 5.53 & 25.89 \\ \hline
  \end{tabular}\label{tab:momstep_benchmark}
}{%
  \caption{Speed-ups obtained for the TGV on the main kernels of the solver}%
}
\end{floatrow}
\end{figure}

\bibliographystyle{ieeetr}
\bibliography{main}

\end{document}